\documentclass[manuscript]{acmart}

\usepackage{array,graphicx}
\usepackage{float}
\usepackage{multirow}
\usepackage{makecell}
\usepackage{soul}
\usepackage{subfig}
\usepackage{graphicx}
\usepackage{CJKutf8}
\usepackage{tikz}
\usepackage{caption}
\usepackage{comment}
\usepackage{hyperref}

\newcommand \redcircle[1]{%
  \tikz[baseline=(X.base)] 
    \node (X) [draw, shape=circle, inner sep=0, fill=red, text=white, minimum width=0.5cm, minimum height=0.5cm] {\strut #1};%
}

\newcommand \blackcircle[1]{%
  \tikz[baseline=(X.base)] 
    \node (X) [draw, shape=circle, inner sep=0, fill=black, text=white, minimum width=0.5cm, minimum height=0.5cm] {\strut #1};%
}

\AtBeginDocument{%
  }

\newcommand{\jieshan}[1]{\textcolor{black}{#1}}
\newcommand{\red}{\textcolor{black}}

\newcommand{\cy}[1]{\textcolor{black}{#1}}

\newcommand{\chyu}[1]{\textcolor{black}{#1}}

\begin{document}

\setcopyright{none}

\title{Prompt Sapper: A LLM-Empowered Production Tool for Building AI Chains}

\author{Yu Cheng}
\email{yc@jxnu.edu.cn}
\orcid{0009-0008-5243-8705}
\affiliation{%
  \institution{Jiangxi Normal University}
  \country{China}
}

\author{Jieshan Chen}
\authornote{Corresponding Author.}
\orcid{0000-0002-2700-7478}
\email{Jieshan.Chen@data61.csiro.au}
\affiliation{%
  \institution{CSIRO's Data61}
  \country{Australia}
}

\author{Qing Huang}
\email{qh@jxnu.edu.cn}
\orcid{0000-0002-8877-4267}
\affiliation{%
 \institution{Jiangxi Normal University}
 \country{China}
}

\author{Zhenchang Xing}
\email{Zhenchang.Xing@data61.csiro.au}
\authornote{Also with Australian National University.}
\orcid{0000-0001-7663-1421}
\affiliation{%
  \institution{CSIRO's Data61}
  \country{Australia}
}

\author{Xiwei Xu}
\email{Xiwei.Xu@data61.csiro.au}
\orcid{0000-0002-2273-1862}
\affiliation{%
  \institution{CSIRO's Data61}
  \country{Australia}
}

\author{Qinghua Lu}
\email{Qinghua.Lu@data61.csiro.au}
\orcid{0000-0002-7783-5183}
\affiliation{%
  \institution{CSIRO's Data61}
  \country{Australia}
}

\renewcommand{\shortauthors}{Cheng et al.}

\begin{abstract}
The emergence of foundation models, such as large language models (LLMs) GPT-4 and text-to-image models DALL-E, has opened up numerous possibilities across various domains. People can now use natural language (i.e. prompts) to communicate with AI to perform tasks. While people can use foundation models through chatbots (e.g., ChatGPT), chat, regardless of the capabilities of the underlying models, is not a production tool for building reusable AI services. APIs like LangChain allow for LLM-based application development but require substantial programming knowledge, thus posing a barrier. 
\cy{To mitigate this, we systematically review, summarise, refine and extend the concept of AI chain by
incorporating the best principles and practices that have been accumulated in software engineering for decades into AI chain engineering, to systematize AI chain engineering methodology.}
We also develop a no-code integrated development environment, \href{https://github.com/YuCheng1106/PromptSapper}{\textcolor{blue}{Prompt Sapper}}, which embodies these AI chain engineering principles and patterns naturally in the process of building AI chains, thereby improving the performance and quality of AI chains. With Prompt Sapper, AI chain engineers can compose prompt-based AI services on top of foundation models through chat-based requirement analysis and visual programming. Our user study evaluated and demonstrated the efficiency and correctness of Prompt Sapper.

\end{abstract}



\maketitle

\section{Introduction}
\label{sec:intro}

The emergence of Large Language Models (LLM) bring many possibilities to how people develop an AI service. In the past, people need to undergo some programming and AI training to collect a large amount of datasets, label them and train a model to achieve an acceptable performance using programming languages. In comparison, foundation models like ChatGPT~\cite{chatgpt}, Dall-E~\cite{ramesh2021zero} relieve these burdens. People can now use natural language (i.e. prompts) to communicate with LLM to distill facts, translate language into another, generate code, revise articles, even obtain professional law advice about their lawsuit~\cite{storkai}. A typical way is that we can simply provide a task instruction like ``Translate this sentence from English to Chinese'', and give a sentence ``Sentence: hello''. The LLM will then understand what we want it to do and then accomplish the task, without having to go through a complex intermediary like a programming language. As a result, it will return ``\begin{CJK*}{UTF8}{gbsn}你好\end{CJK*}'', which is ``hello'' in Chinese.

While the current chatbot-based interface, like ChatGPT playground~\cite{chatgptPlayground}, enables non-technical users to encounter and use the state-of-the-art powerful LLMs to simplify and accelerate their daily life using natural language, it suffers from reusability issues.
The user needs to repeatedly have the same conversations with chatbot everytime they want to perform a task. This problem is not obvious in simple tasks, but get quite annoying and time-consuming when it comes to complex tasks that may involves over three rounds of conversations. The problem aggravates when the task involves different LLMs.
While package like LangChain~\cite{langchain} provides well support to deal with above issues, it requires programming background, which creating a natural barrier for non-technical users.
An infrastructure that tackle these issues while requiring no/little programming skills are needed.




On the other hand, the concept of \textit{AI chains} emerges over the time~\cite{levine2022standing, huang2023pcr, wu2022ai, wu2022promptchainer, schick2022peer, yang2022re3, arora2022ask, creswell2022selection, singh2022progprompt, creswell2022faithful,jiang2022promptmaker,huang2023ai}.
AI chains involve the creation of complex and specialized AI services by sequentially chaining prompt-based AI functionalities on foundation models, following a predefined workflow. This approach requires minimal or no coding in traditional programming languages.
This ability is aligned with human way to deal with complex tasks. 
Some researchers have attempted to lower the barriers to prototyping AI functionality through visual programming~\cite{jiang2022promptmaker, wu2022promptchainer, wu2022ai}, but they do not address the challenges of task decomposition, prompt design, evaluation, and management when creating AI functionality based on foundation models. 
Meantime, other have developed some scattered tools~\cite{aibuilder, superbioai,structuredprompt}, but these tools are independent and cannot systematically support AI chain methodology. 

In this paper, we review existing AI chain work and introduce a comprehensive AI-chain methodology to systematize prompt engineering practices, with the aim of advancing the modularity, composability, debuggability and reusability of AI functionalities.
It consists of three basic concepts of AI chains, namely, AI chain requirements, worker composition and cooperation and worker roles and prompts.
We also introduce a ``magic-enhancing-magic'' that utilise LLMs to assist the development process.
Finally, we systematically define and discuss the three activities (AI chain system design, AI chain implementation and AI chain testing) embedded in the AI chain system design phases.

Incorporating our methodology, we develop an AI chain integrated development environment (IDE), Prompt Sapper~\cite{promptsapper}, to support the full life cycle of AI chain development. By organically embodying AI chain methodology in this IDE, people can naturally apply AI chain best practices throughout the AI chain development process, enabling those with limited programming skills to develop high-quality foundation model-based AI services.
The AI services can be deployed as web services to provide services to end users through websites or mobile applications.
They can also be embedded in large software projects or other design and development environments (e.g, Figma) to facilitate rapid AI functionality prototyping.
Therefore, Prompt Sapper can democratize the usage of LLMs, making AI accessible to a wider range of individuals, including non-technical people.

To evaluate the usefulness of Prompt Sapper, we conducted a user study with \cy{18} participants. The results demonstrates the low entry barrier and practicality of our Prompt Sapper. It can significantly save time for AI chain engineers compared to using traditional IDEs for programming by alleviating the need to learn various API calls and reducing the time spent searching for online resources. This intuitive and accessible approach serves to democratize the usage of foundation models. 
\cy{Additionally, we conducted another user study with 12 participants to investigate the effectiveness of the two LLM-based co-pilots within Prompt Sapper. The results showed that these co-pilots enhance the efficiency and effectiveness of the design phase of AI chain development, further enhancing the overall utility of our system.}



In summary, in this paper, we make the following contributions: 
\begin{itemize}
    \item We \red{reviewed, summarised, refined and extended the existing work on AI chains and introduce} a comprehensive AI chain methodology that systematizes the best principles and prompt engineering practices.
    \item We developed a block-based visual programming tool, Prompt Sapper~\cite{promptsapper}, an AI chain infrastructure, embedding our methodology and LLM co-pilots to enable non-technical users to properly and seamlessly develop their own LLM-based AI chain services in a natural way. Furthermore, we released the source code of our tool at  \href{https://github.com/YuCheng1106/PromptSapper}{our GitHub repository}~\footnotemark[1], making it accessible to the wider community.
    \item \cy{We conducted three user studies\footnotemark[1] with a total of 30 participants, collectively demonstrating the usefulness, efficiency, correctness, and effectiveness of the proposed tool and its two LLM-based co-pilots in the AI chain generation process.}
\end{itemize}

\footnotetext[1]{
\cy{The implementation of Sapper IDE and all experiment data and results can be downloaded at our Github repository: \url{https://github.com/YuCheng1106/PromptSapper}}
}











\section{Background and Related Work}
In this section, we will introduce the concepts and evolution of large language models and existing visual programming tools.

\subsection{Large Language Models}

Generative AI models, such as those represented by the GPT series~\cite{radford2019language, brown2020language, openai2023gpt4}, DALL-E~\cite{ramesh2021zero}, Stable Diffusion~\cite{rombach2022high}, and MidJourney~\cite{midjourney}, are rapidly developing and are often referred to as foundation models~\cite{bommasani2021opportunities}. These models exhibit great capabilities for enabling different tasks without requiring any fine-tuning, ranging from context rewriting, code generation, to design generation. Researchers have focused their efforts on using prompt-based methods to enable these foundation models to perform various tasks~\cite{huang2023pcr, schick2022peer, yang2022re3, creswell2022selection, kazemi2022lambada, arora2022ask,singh2022progprompt,huang2023ai}.
The GPT3Demo website~\cite{gpt3demo} lists over 800 GPT-based apps and use cases. With the help of natural language prompts, users can communicate their task requirements to these models, which can efficiently and accurately perform the task and produce acceptable answers. However, prompt engineering is a non-trivial task, and many prompting tricks and patterns have been proposed to aid in the process.

As researchers explore the potential of Language Models (LLMs), they have discovered a misalignment between human-crafted prompts and what LLMs have learned from their datasets, resulting in unreliability, instability, and poor performance in some tasks. 
To fully leverage the capabilities of LLMs, researchers have worked out different strategies to overcome this challenge~\cite{brown2020language, zhao2021calibrate,madaan2022language,ye2023context}.
Brown et al.\cite{brown2020language} propose using in-context few-shot learning to let LLMs implicitly learn more specific requirements. 
Ye et al.\cite{ye2023context} propose to better clarifying the requirements by adding the task input and output.

When it comes to reasoning tasks, a direct request to provide an answer is not sufficient. Kojima et al.\cite{kojima2022large} found that by simply adding the magic sentence ``let's think step by step'' \red{to the end of the prompt, GPT-3 can produce multi-step reasoning in a zero-shot setting and finally come out with the correct answer.} This results in great improved performance in reasoning tasks. 
Wei et al.\cite{wei2022chain} later propose chain-of-thought prompting by generating a series of intermediate reasoning steps to enhance LLM performance in complex reasoning.
There are also other strategies to improve prompts gradually, such as self-ask~\cite{press2022measuring}, self-reflection~\cite{shinn2023reflexion}, and self-consistency~\cite{wang2022self}.

\red{Later on, researchers gradually find that while a one-round conversation with LLM may work well for simple tasks, more intricate tasks that involve multiple steps and nuances requires extra considerations, without which the output of LLM may deviate from expectations, leading to suboptimal or unsatisfactory responses.}
As a result, the idea of AI chains has emerged. Rather than asking LLMs to split tasks or think aloud the process, people can be involved in the loop to break down tasks into smaller, simpler ones, and then ask the model to summarize the output of each module. PromptChainer, proposed by Wu et al.~\cite{wu2022promptchainer}, enables users to break down a complex task into several simple tasks using flow-chart based visual programming, and then chain these tasks and prompts together.
Their concurrent work, AI chains~\cite{wu2022ai}, further demonstrates the necessity of chaining prompts instead of relying on one-round prompts, which largely enhances the quality of response and promotes transparency, controllability, and collaboration.
Levine et al.~\cite{levine2022standing} proposed using multiple calls on frozen LLMs to complete complex tasks, and a large amount of research has found that task decomposition can enhance the reasoning ability of foundation models and enable them to complete more challenging problems.
However, existing attempts are either task-specific, limited to a single LLM, or lack of flexibility. 
\red{ChatGPT Plugins~\cite{chatgptplugin} empower GPT to seamlessly integrate with other plugins or models, providing real-time access to information, computational capabilities, and third-party services (like BrowserOp for browsing webpages, Code Runner for immediate code execution). It treats LLMs as central controllers, planning and executing the workflow autonomously. However, the inherent unpredictability of the model means outcomes can vary with each run, potentially compromising its repeatability and consistency, and the users thus lose control over the models.
}
Moreover, a systematic AI chain methodology is missing in the new era of foundation models.


Our work aims to systematize the AI chain methodology by reviewing a comprehensive set of papers and combining our own experience to establish best principles and practices. Our objective is to develop a block-based visual programming tool, called Prompt Sapper, which incorporates our methodology and LLM co-pilots, allowing users to inadvertently learn the methodology while programming. Our tool supports various workers ranging from non-AI workers to the latest AI 2.0 models (LLMs) and Software 3.0 and facilitates collaboration between different workers and models. We prioritize a human-centric design, ensuring that users can easily comprehend how the integrated development environment (IDE) functions. Furthermore, we offer users the freedom to download their created AI service and deploy it as a plugin in their working environment.
\red{Moreover, similar to ChatGPT Plugin, Our Prompt Sapper also incorporates a comparable plugin mechanism, fostering collaboration between various models via the engine management module. Unlike the ChatGPT Plugin, it sidesteps usability challenges. Users maintain control over the AI chain's construction, ensuring a transparent and predictable process through standard programming constructs such as input, output, if-else statements, variables, and loops.}





\subsection{Visual Programming Tools}
%
Many tools have been proposed to support various tasks using no code and low code solutions, spanning different fields, like website generation~\cite{wix}, robot-IoT workflow~\cite{huang2020vipo}. 
There are mainly two types of visual programming, i.e., form-filling~\cite{ifttt} and visual programming languages~\cite{wu2022ai}. 
Form-filling enables users to fill in a form to add actions and conditions. 
For example, IFTTT (if this then that)~\cite{ifttt} allows users to quickly create an applet by filling the forms to define a trigger point and an action. The website provides many predefined trigger points of many apps, like ``new tweet by you with hashtag'' for Twitter, and actions like ``add a row to spreadsheet'', or some AI functions like summarising the article.
Steinmetz et al.~\cite{steinmetz2018razer} propose Razer, which allows non-expert people to automate production processes by defining action sequences of robots via form filling method.
\cy{Jiang et al.\cite{jiang2022promptmaker} develop PromptMaker that expedites the prototyping of new machine learning (ML) functionality through natural language prompts. It offers both freeform and structured prompt creation modes, and the users can get the response from the models (LLMs) instantly so that they can revise their prompt to fit their requirement.}
Their research found that prompt programming can effectively speed up the prototyping process and promote communication between collaborators, highlighting the need for tool support for designers.
However, these works are idiosyncratic and application-specific.

On the other hand, visual programming provides visual constructs like conditions to support more flexible no code or low code methods. 
Wu et al.~\cite{wu2022ai} propose a flow-chart like visual programming method to allow users to break down complex tasks into several simple tasks, and rely on a single LLM to do some small tasks and combine the results.
However, their tool can only work on linear/parallel workflow, and does not support loops, variables and multiple LLM collaborations. It is also hard to collaborate with different workers.

Our AI chain methodology and IDE support the no/low-code development of AI services based on foundation models. 
Our methodology and IDE are fundamentally different from existing no-code AI tools. 
Many no-code AI tools, such as Apple ShortCut, IFTTT~\cite{ifttt}, Zaiper, Microsoft Power APP, and Azure AI Builder, are designed for robotic process automation (RPA), which \red{means the use of software robots or bots to automate repetitive and rule-based tasks within business processes}.
They supports different business scenarios by integrating multiple applications used in a workflow. 
These tools incorporate pre-built AI models to perform data analysis tasks, such as determining email sentiment \red{to auto sort emails into different folders, forward to specific individual or trigger certain notifications}. 
Therefore, these tools focus on utilizing AI services in a no-code way, rather than developing these AI services. 
In contrast, our AI chain IDE aims to support AI service development, which can be embedded into business processes using these RPA tools.

\section{AI Chain Methodology and IDE requirements}

In this section, we review existing AI chain works like~\cite{levine2022standing, huang2023pcr, wu2022ai, wu2022promptchainer, schick2022peer, yang2022re3, arora2022ask, creswell2022selection, singh2022progprompt, creswell2022faithful,jiang2022promptmaker,huang2023ai} and define a comprehensive AI chain methodology to systematize the AI chains engineering best principles and practices. We first introduce the basic concepts of AI chain, and then introduce ``magic enhancing magic'' activity that can enhance AI chain engineering. 
Finally, we describe three key activities from AI Chain System Design, AI Chain Implementation to AI Chain Testing.

\subsection{Concepts of AI Chain}
AI chains engineering inherits and adapts traditional software concepts such as software requirements, and object composition and collaboration, but also develops unique concepts specific to AI chains such as prompt designs and patterns.
These concepts have different level of relevance to different AI chain engineering activities, and they will be identified, analyzed and refined in various AI chain engineering activities throughout the rapid prototyping process of AI services.

\subsubsection{AI Chain requirements} 
Building AI chains is a rapid prototyping process that creates custom AI services on top of foundation models.
Foundation models make rapid prototyping of AI services feasible, as we no longer need to spend time and effort on data engineering and model training, but focus on the act of solving problems using AI.
Furthermore, rapid prototyping allows for the quick delivery of working software and obtaining feedback to iteratively improve AI services.
As AI chain engineering frees us from low-level coding, we will see a revival of important but undervalued software engineering activities, such as requirement analysis, specification and verification~\cite{automaticProgramGeneration} throughout AI chain engineering. 
The key is to align task needs and problem understanding with model capability through iterative design, evaluation and improvement of AI chains and prompts.
This process can be enhanced by the interaction with large language models, which we call ``magic enhancing magic'' in Section~\ref{sec:magicenhancingmagic}.

\subsubsection{Worker composition and cooperation}
An AI chain consist of a set of cooperating function units, which we call workers, analogous to objects. 
AI chain is a recursive concept.
Depending on the task complexity and model capability, an AI chain may organize cooperating workers into a hierarchy of workers. 
The worker hierarchy can be represented in composite pattern~\cite{gamma1995design}, in which the leaf worker has no children and implements the functionality directly, and the composite worker (analogous to module) maintains a container of child workers (leaf or composite) and forwards requests to these children.
An AI chain also needs to specify the workflow, control structure, and cooperation of its workers.
Workers need to define a ``function signature'' for communicating with human and other workers.
We can use workflow patterns~\cite{workflowpatterns} to define and implement cooperation between workers, data management, and exception handling.
These worker design aspects are analogous to system design and should follow computational thinking principles, such as problem decomposition, pattern recognition, and algorithmic thinking, as well as well-recognized software engineering practices, such as separation of concerns and modularity.

\subsubsection{Worker roles and prompts}
A prompt defines the function of a worker and ``program'' the foundation model to complete the corresponding task.
The workers should follow single responsibility principle and play distinct roles instructed by their prompts.

While prompt programming lowers the barrier for non-technical individuals to develop AI prototypes, the difficulty of designing effective prompts increases with increasing reasoning ability of the worker.
Some challenges arise in finding or generating representative few-shot examples (so called example sourcing~\cite{jiang2022promptmaker}), requiring AI chain engineers to be creative and maintain example datasets (although would be smaller than those typically used to train neural networks). 
Another challenge is to accurately describe task workflows, which can be achieved using semi-structured or code-like prompts~\cite{singh2022progprompt, madaan2022language, mishra2021reframing}. 

By decomposing complex workflows into cooperating workers, it significantly reduces prompt design difficulty as each worker only needs to execute a simple sub-step of the complex workflow.
Furthermore, we can provide each sub-task worker with sufficient specific examples, test and debug individual workers (akin to unit testing) to optimize its performance, and seamlessly plug the improved worker into the AI chain, as a systematic way to improve performance on the complex task.
A well-design worker can be reused across multiple tasks.

\subsection{Magic Enhancing Magic}
\label{sec:magicenhancingmagic}

This co-pilot activity leverages the capabilities of LLMs to enhance AI chain engineering from two aspects: requirement elicitation and mechanical sympathy.
Our Sapper IDE is equipped with specialized LLM-based co-pilots to support the magic-enhancing-magic activity in different phases of AI chain engineering. 

\subsubsection{Requirement elicitation}
Like other software projects, AI chain engineers often start with a vague understanding of ``what they want''.
Building software on such vague statements can be a major reason for the failure of software projects.
We need to interact with engineers through requirement elicitation and gradually clarify the initial ``what they want'' statements into specific AI chain needs. 
This challenging task can be supported by large language models. 
By providing some examples of requirement elicitation, such as good open-ended requirement elicitation questions in the Software Requirements textbook~\cite{wiegers2013software} by Karl E. Wiegers, the LLMs would learn to ask good open-ended questions for specific tasks and thus elicit specific AI chain needs.
The Design view of our Sapper IDE (see Section~\ref{sec:designview}) is equipped with such a requirement elicitation co-pilot that interacts with the engineer to elicit and analyze tasks requirements.

\subsubsection{Mechanical sympathy}
Mechanical Sympathy means that a racing driver does not need to be a mechanical engineer, but understanding how the car works can make one a better racing driver. 
The same applies to developing AI chains and writing prompts, especially as large language models are transforming AI from traditional design and engineering to something more akin to natural science~\cite{AIEns}. 
Therefore, we need to empirically explore and understand emergent AI behaviors and capabilities. 

Gwern Branwen's blog~\cite{chatgptPaP} proposes that we need to anthropomorphize prompts.
We need to consider how people would typically express the information we need if they had already discussed it on the Internet. 
Preliminary research~\cite{gonen2022demystifying} suggests that the more familiar the model is with the language used in the prompt, the better the prompt's performance tends to be.
We also need to test different prompts to see what different outputs they may lead to and reverse engineer how large language models understand these prompts, thereby discovering any discrepancies between what we assume/expect and what the models understand.
Understanding these discrepancies can guide us in decomposing tasks and writing prompts to fit with the model capability.

Foundation models provide playgrounds for experimenting prompts.
Research has shown that large language models can create human-level prompts~\cite{zhou2023large}.
Inspired by this, our IDE's Design view (Section~\ref{sec:designview}) implements a ChatGPT-based prompting co-pilot to create candidate prompts for the generated AI chain steps as initial task decomposition (Section~\ref{sec:taskdecomposition}).

\subsection{AI Chain System Design}
\label{sec:systemdesign}
The AI chain design and construction phases are highly iterative, and the AI chain system design will be iteratively optimized based on the actual running results of the AI chain.

This activity is a continuation and a starting point: a continuation in the sense that it clarifies and refines the initial (vague) ``what we want'' and the approximate task model into specific AI chain requirements through requirement analysis, and a starting point in the sense that it produces the AI chain skeleton as the foundation for implementing the AI chain through task decomposition and workflow walk-through. These two activities adopt the basic principles of computational thinking and the modular design of software engineering, and will determine worker composition, forming a skeleton of AI chain.

\subsubsection{Requirement Analysis}
\label{sec:requirementanalysis}

The initial AI chain requirements provided by AI chain engineers are often incomplete or ambiguous. Building the AI chain on such requirements makes it difficult to ``build the right AI chain''. In software projects, it is usually necessary to clarify and refine requirements through communication between product managers and users. We achieve this through an LLM-powered requirement co-pilot, which leverages its vast knowledge to iteratively rewrites and expands the AI chain requirements.

\subsubsection{Task decomposition}
\label{sec:taskdecomposition}
When given a complex multiplication, such as 67*56, many people may not be able to provide the answer immediately. 
However, this does not mean that humans lack the ability to perform two-digit multiplication. 
With a little time, pen, and paper, it is not too difficult to stepwise calculate 60*50+7*50+60*6+7*6 and arrive at the final answer. 
The same principle applies when working with foundation models. 
When a model cannot solve a problem successfully in a generative pass, we could consider how humans would divide and conquer such complex problems. 
Consulting large language models can often provide inspiration for task decomposition.

We can iteratively break down the problem into small, simpler ones.
If a sub-task is still too complex, it can be further decomposed into simpler sub-tasks until each small problem can be handled by the model in a single generative pass. 
This results in a hierarchy of sub-tasks, each of which is accomplished by an AI chain worker. 
We achieve this through an LLM-powered decomposition co-pilot, which facilitates the process of breaking down complex tasks into manageable sub-tasks. 

\subsubsection{Workflow walk-through}
AI chain workers collaborate to complete tasks according to a workflow.
This requires algorithmic thinking to develop a step-by-step process for solving problems.

The AI chain workflow needs to support control structures, such as condition and loop.
For example, in order to avoid introducing errors into correct code, PCR-Chain~\cite{huang2023pcr} only attempts to fix the code when the LLM determines that there are last-mile errors (judging whether the code has errors is easier than fixing the code). 
Re3~\cite{yang2022re3} works by iteratively expanding a story based on an outline, while maintaining coherence and consistency throughout the story.
Meanwhile, each worker needs to define a function signature in terms of input and output as well as any pre/post-condition constraints, akin to interface specification in software engineering, so that workers can interact or communicate with one another.

\subsection{AI Chain Implementation}
\label{sec:aichainimplementation}


\textit{Grice's maxims of conversion:}
As a means of conversion, prompts should obey Grice's maxims of conversion~\cite{CoopPrin}: informative, truth, relevance, and clarity. 
From the computational thinking perspective, it is about identifying the essential elements of a problem and removing unnecessary details.
One should avoid including irrelevant details or unrelated statements, as in human-written text, such details are typically assumed to be relevant, regardless of how illogical a narrative incorporating them may appear~\cite{chatgptPaP}.

\subsection{AI Chain Testing}
\label{sec:aichaintesting}

AI chain testing involves running an AI chain and its component workers (prompts), evaluating the running results, and debugging and fixing unexpected behaviors.
Similar to traditional testing, we can perform unit tests for each worker, integration (interface) tests for cooperating workers, and system tests for the entire AI chain. 
Our IDE's Block view supports all three levels of testing.

Since the behavior of foundation models is un-designed but emergent, testing methods based on verifying design and construction correctness are no longer applicable. 
AI chain and prompt testing are more of an experimental process to gain mechanical sympathy for the model and to find effective prompts to interact with the model.
In our IDE, users can test and debug a worker's prompt in the Block view.

\section{Prompt Sapper: AI Chain IDE}
\label{sec:sapperide}



Unlike traditional IDEs focused on code development, considering that AI chains will be developed by a large number of people with little or no computing knowledge or skills, the highest goal of our AI chain IDE design is to put people first. 
This is reflected in three design principles. 
First, we seamlessly embody promptmanship (AI chain methodology) in the AI chain IDE, effectively utilizing AI chain methodology and best practices in anyone's AI chain development process. Second, we make full use of the LLM's extensive knowledge and conversational capabilities, turning it into intelligent co-pilots for AI chain development, effectively assisting non-technical professionals in completing AI chain task modeling, requirement analysis, design and construction. 
Third, we support the whole-process no-code development and deployment of AI chains. 
To that end, we explicitly support the exploration and design phases in the IDE, providing effective support for AI chain analysis and design, enabling non-technical people to complete the main work of AI chain development in a natural, non-programming way, thus minimizing what needs to be done in AI chain implementation. 
Furthermore, we use blocky-style visual programming to support AI chain implementation, execution, and testing, further minimizing the programming skills required. 
Finally, through one-click deployment, anyone can easily complete the last mile between AI services and end users.

\begin{figure*}
    \centering
	\includegraphics[width=\textwidth]{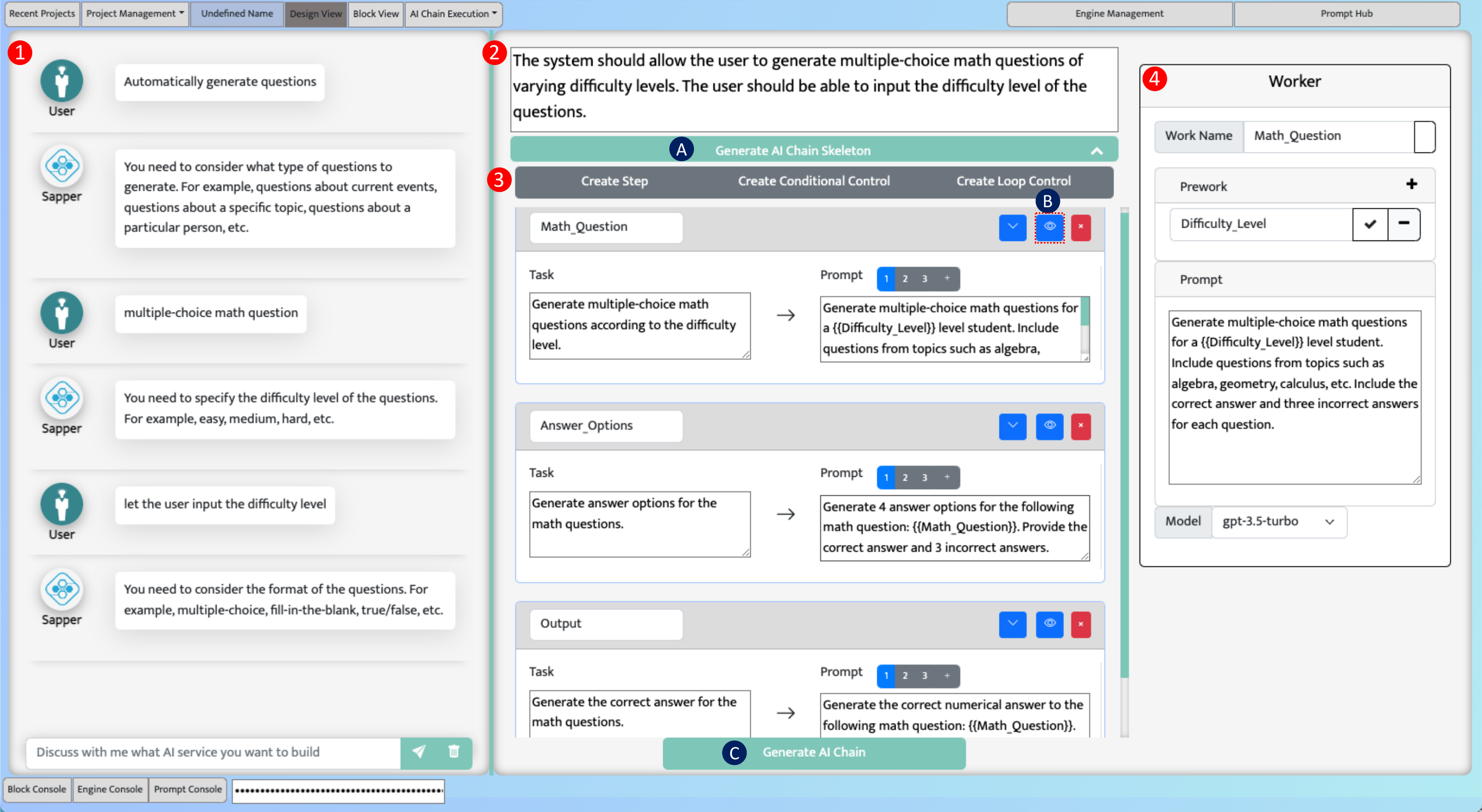}
	\caption{Design View. The user first consults our co-pilot to elicit their requirements \protect\redcircle{1}, which are later summarised in  \protect\redcircle{2} by another co-pilot. Once the user's task description is specific enough, they then click \protect\blackcircle{A} to generate the skeleton \protect\redcircle{3}. They can modify the skeleton, adding/deleting control in \protect\redcircle{3}, and click \protect\blackcircle{B} to modify input, prompts and models in  \protect\redcircle{4}. Once the user finishes the modification, they can click \protect\blackcircle{C} to generate the block code in Block View (Fig.~\ref{fig:block_view})
 }
	\label{fig:design_view}
\end{figure*}

\subsection{Design View}
\label{sec:designview}
Design view supports the main activities of the design phase and plays an important role in bridging the gap between the exploration and construction phases. 
Therefore, it has two main functions: requirement elicitation and AI chain skeleton generation, supported by two LLM-based co-pilots.
\cy{Additionally, the primary objective of the two co-pilots in the Design view is not solely to directly minimize errors during AI chain construction, but rather to actively interact with users, fostering requirement elicitation and AI chain skeleton generation.}

Fig.~\ref{fig:design_view} shows the usage of Design View.
On the left side, there is a LLM-based chatbot (Fig.~\ref{fig:design_view}~\redcircle{1}). 
Design view's chatbot is prompted as an infinite questioner, which utilizes the vast knowledge of a LLM to generate clarifying questions based on a user's initial input to support requirement elicitation and analysis.

The conversation starts with the user's task description (usually a vague description of what the user wants).
Based on this initial task description, the Design view's chatbot asks a series of clarification questions to elicit the user's specific task requirements. 
Each round of clarification response from the user is iteratively incorporated into the user's task description (displayed in the Task Description box Fig.~\ref{fig:design_view}~\redcircle{2}). 
For example, if the user's initial task description is ``Automatically generate questions'', and the responses to two rounds of clarifying questions are ``multiple-choice math questions'' and ``let the user input the difficulty level'', the task description will be updated to ``The system should allow the user to generate multiple-choice math questions of varying difficulty levels. The user should be able to input the difficulty level of the questions.'' (Fig.~\ref{fig:design_view}~\redcircle{2}).
Of course, if the user believes he already has clear requirement and thus does not need the help of the requirement analysis co-pilot, they can directly enter the requirement in the task requirement box.

At any time (e.g., when the user feels the task description is specific enough), the user can click the ``Generate AI Chain Skeleton'' button (Fig.~\ref{fig:design_view}~\blackcircle{A}) below the Task Description box to request the Design view to generate the main steps required to complete the task, as well as three candidate prompts for each step. 
To achieve this, Design View implements another LLM-based co-pilot. 
This co-pilot first converts the high-level intention in the task description into the main steps required to complete the task.
For example, for the task description in the Task Description box, it generates three steps (Fig.~\ref{fig:design_view}~\redcircle{3}): ``Generate multiple-choice math questions according to the difficulty level'', ``Generate answer options for the math questions'' and ``Generate the correct answer for the math questions''. 
Each step has a step name (used to identify the step output in the AI chain) and a concise step description. 
Then, based on the step description, the co-pilot recommends three candidate prompts for each step. 
For example, for step ``Generate multiple-choice math questions according to the difficulty level'', a generated prompt is ``Generate multiple-choice math questions for a \{\{Difficulty\_Level\}\} level student. Include questions from topics such as algebra, geometry, calculus, etc. Include the correct answer and three incorrect answers for each question''. 
The user can also modify the prompt or directly enter their own one.

The user can manually modify the generated steps, including remove steps, add steps, or reorder steps (Fig.~\ref{fig:design_view}~\redcircle{3}). 
In the current version, the co-pilot does not generate control flow. 
The user can manually add control flow (such as step execution conditions or step loops) or add control flow during visual programming of AI chain. 

The user can edit the generated prompts using a structured form (Fig.~\ref{fig:design_view}~\redcircle{4}) by expanding the task via (Fig.~\ref{fig:design_view}~\blackcircle{B}). In this form, the user can also set the input to the step and select the engine to execute the prompt.

\begin{figure*}
    \centering
	\includegraphics[width=\textwidth]{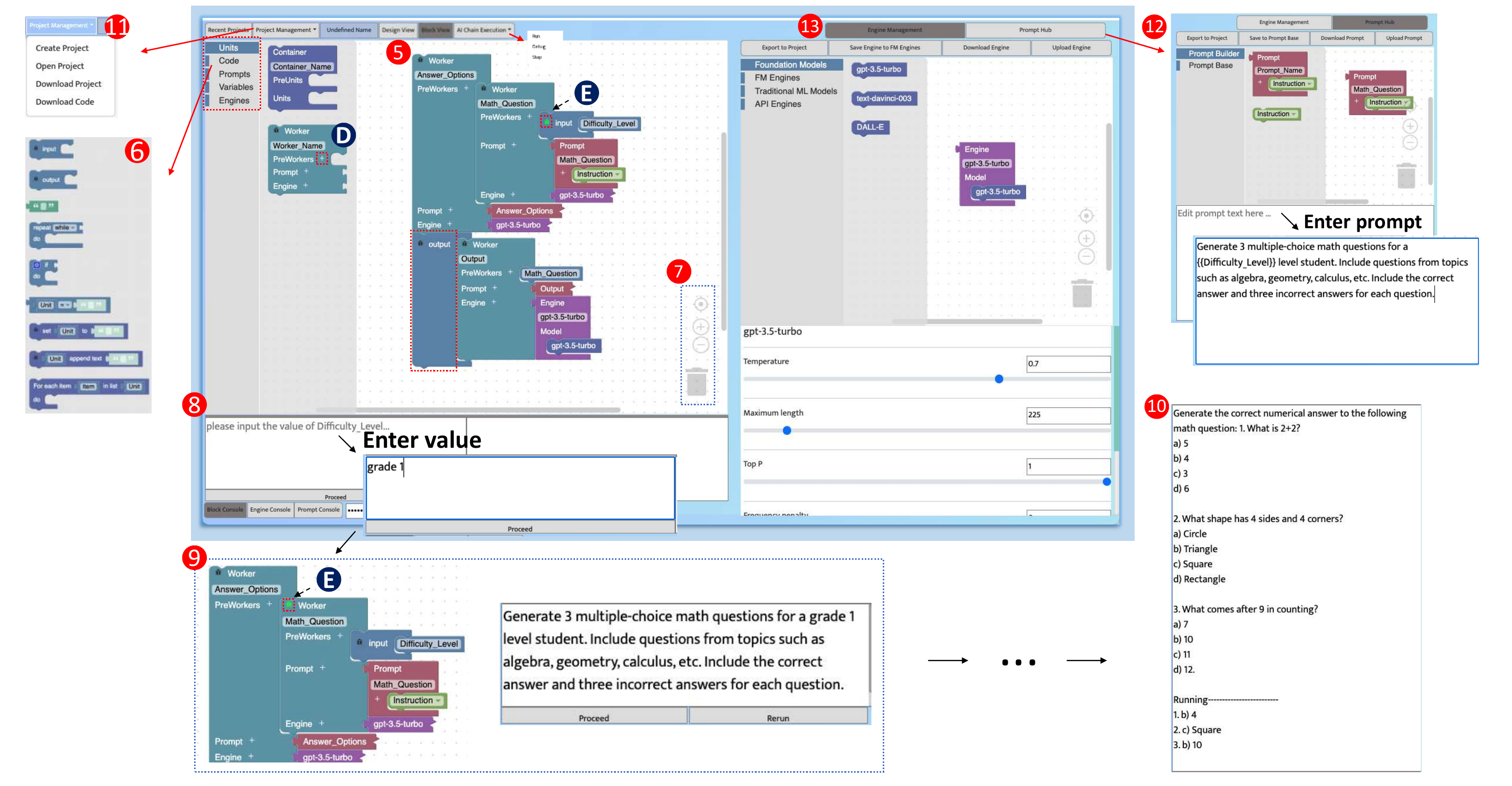}
	\caption{\cy{Block View and Artifact Management.}}
	\label{fig:block_view}
\end{figure*}

Finally, the user can click the Generate AI Chain button (Fig.~\ref{fig:design_view}~\blackcircle{C}) at the bottom right of Design view, and the IDE will automatically assemble a block-based AI chain according to the AI chain skeleton, which can be viewed, edited, and executed in the Block View (see Section~\ref{sec:blockview} and Fig.~\ref{fig:block_view}).

\subsection{Block-based Visual Programming}
\label{sec:blockview}

We use block-based visual programming to support AI chain implementation, execution, and debugging. 
The current implementation is based on the open-source Blockly project~\cite{blockly}.
We show a usage example in Fig.~\ref{fig:block_view}.

\textbf{Worker Block:}
The core component of an AI chain is the worker. 
A \textbf{worker block} consists of three slots: preworkers, prompt, and engine.
A new Worker block can be dragged to the editor from the Units toolbox.
The \textit{preworkers} slot specifies a worker's inputs, which can be zero or more Input or Worker blocks. 
An input block receives (from the Code toolbox) user input from the console or uses a variable as input to the worker. Connecting a worker A to another worker B's prework slot is equivalent to running these two workers in series, using A's output as B's input. 
The \textit{prompt} slot can hold a natural language prompt. Prompt can be in the form of plain text, semi-structured text or code-like text. The prompt slot can also hold a piece of executable code (which can be thought of as a special type of prompt).
The \textit{engine} slot specifies the foundation model used to execute the prompt or the tool used to run the code (such as a Python interpreter). 

\textbf{Container Block:}
We design \textbf{container block} to represent composite workers (see Fig.~\ref{fig:block_view}~\redcircle{5}). 
A new Container block can be dragged to the editor from the Units toolbox.
A container block has two slots: preunits and units. 
The \textit{preunits} slot specifies a composite worker's inputs, which can be zero or more workers (leaf or composite). These preunits will be executed before the present composite worker.
The \textit{units} slot can hold one or more worker or container blocks, forming a worker hierarchy, as well as any traditional code blocks for user input and control structure.

\textbf{Code Block:}
The Code tab includes traditional programming constructs such as console input, console output, assignment, if, for, while blocks (Fig.~\ref{fig:block_view}~\redcircle{6}).
Users can drag and drop the required code blocks into the editor and assemble them with the worker/container blocks by dragging and dropping. If a worker block needs to output information to end-users, it need to be placed in an Output block (see the lower part of Fig.~\ref{fig:block_view}~\redcircle{5}). Then, the output of this worker will be displayed in the bottom-right Output window. Otherwise, the worker's output can only be inspected in the Block Console during the worker execution, but not visible to the end users.

\textbf{Prompt, Variable and Engine blocks:}
Prompts, variables and engines used by the worker blocks are managed as explicit blocks in the corresponding toolbox. In the Prompts and Engines toolbox, the user can create/import prompts from Prompt Hub (Fig.~\ref{fig:block_view}~\redcircle{12}) and engines managed in Engine Management (Fig.~\ref{fig:block_view}~\redcircle{13}). 
In the Code toolbox, the user can create or delete variables.

\textbf{Edit worker/container blocks:}
To make it intuitive for people to build and modify AI chains, all visual programming operations can be triggered directly on the module/worker unit. 
Clicking the + (Fig.~\ref{fig:block_view}~\blackcircle{D}) on the right side of each slot can directly add or edit the corresponding blocks. 

\textbf{Edit AI chain:}
A block can be added to the editor by dragging and dropping a block template from the toolbox. The use can assemble the blocks by dragging and dropping in the editor. The user can zoom in/out the editor or place the selected block in the center of the editor by clicking the ``+'', ``-'', and ``aim'' button at the right side of the editor (Fig.~\ref{fig:block_view}~\redcircle{7}).

Selecting a block and pressing delete can delete a block. Deleted blocks can be restored from the Recycle Bin at the bottom-right corner of the editor. By right-clicking a block, the user can duplicate, delete, collapse, or disable a block from the context menu. A Worker and Container block can be shown in a compact form to save space (by Inline Inputs from the context menu).
Collapsing a block will show the worker as a node. This saves the space for other blocks and is helpful when viewing and editing complex AI chains. A collapsed block can be restored by expanding the collapsed node.
Disabling a block will exclude the block from the AI chain execution. This is useful when testing variant workers in an AI chain. A disabled block has the gray grid background, and can be enabled back to the normal status.
The user can right-click a block and add a comment which will be accessed and edited by the "?" icon at the top-left corner of a block.

\textbf{Run and debug AI chain and workers:}
Users can run or debug the AI chain . 
When a worker is running, the bug signal in the top-left corner of the worker block will light up (Fig.~\ref{fig:block_view}~\blackcircle{E}). The actual prompt used during the worker's execution and the engine outputs will be outputted to the Block Console. User inputs required for execution will be entered in the Block Console.

In debug mode, workers will be executed one at a time (Fig.~\ref{fig:block_view}~\redcircle{5} -> \redcircle{8} -> \redcircle{9} -> \redcircle{10})). After a worker finishes running, the execution will be suspended, and the user can check if the outputs in the Block Console match their expectation (Fig.~\ref{fig:block_view}~\redcircle{8}). If the result is as expected, the user can proceed to execute the next worker. Alternatively, the user can modify the prompt for the current worker in the Prompt Console and then re-run the current worker.

If a worker block is placed in an Output block, its output will be displayed in the bottom-right Output window. This window does not show the output of workers that are not placed in an Output block, nor will it show prompts.
The Block Console is used to help AI chain engineers debug the AI chain and thus contains prompt information and intermediate execution results. The Output Window at the bottom-right allows the engineer to inspect the AI chain output that end-users will see (Fig.~\ref{fig:block_view}~\redcircle{10}).

\subsection{Artifacts Management}
\label{sec:artifactsmanagement}

AI chain development produces and uses many different software artifacts that need to be effectively managed. 
Following the principle of separation of concerns, we divide these artifacts into three parts for management: project (Fig.~\ref{fig:block_view}~\redcircle{11}), prompt (Fig.~\ref{fig:block_view}~\redcircle{12}), and engine (Fig.~\ref{fig:block_view}~\redcircle{13}).

\subsubsection{Project Management}
Users can create a new AI chain project (Create Project), download the current project to local disk (Download Project), or open a project on local disk in the IDE (Open Project).
Users can download the backend Python code implementing the AI chain (Download Code) to local disk and reuse the code in their other software projects. Note that our sapperchain Python library (which is not open sourced) is required to execute the downloaded AI chain code.

\subsubsection{Prompt Hub}
Similar to the value of code in traditional software, prompts are the core assets in the AI chain paradigm, which can be reused across AI chain projects. Therefore, our Sapper IDE supports a Prompt Hub (Fig.~\ref{fig:block_view}~\redcircle{12}) that is independent of project management. Prompt Hub has two tabs: Prompt Builder and Prompt Base.

In the Prompt Builder tab, the user can manually create new prompts. They can also drag an existing prompt from the Prompt Base tab to the prompt editor to edit it. Editing prompt blocks in the prompt editor is the same as editing other types of blocks in the AI chain editor.
Experienced users can directly enter or edit prompt text in the Prompt Text view at the bottom. Users can also create or edit prompts in a structured way by the four prompt aspects (Context, Instruction, Examples, and Output Formatter). Except for the Instruction, the other three aspects are optional, but are recommended to represent the prompt in a more structure form.
Clicking "Save to Prompt Base" will save the created prompts to the Prompt Base or update existing prompts in the Prompt Base. All prompts can be viewed in the Prompt Base tab. 

In the Prompts tab of an AI chain project, the user clicks the ``Import Prompt...'' button, which will open the Prompt Hub view. The user can then drag the prompts that the project will use from the Prompt Base tab to the prompt editor. Then, clicking the ``Export to Project'' button will export the selected prompts to the project. The user can change prompts used in the project in the project's Prompt Console tab, but these changes will not affect the original prompts in the Prompt Hub.
The user can download the prompts into local files or upload local prompt files to the IDE.

\subsubsection{Engine Management:}
Workers use different types of engines to complete tasks, such as foundation models and external APIs. These engines can also be used across AI chain projects, so the AI chain IDE supports separate Engine Management (Fig.~\ref{fig:block_view}~\redcircle{13}).
Our IDE currently pre-installs three foundation models, including gpt-3.5-turbo, text-davinci-003, and DALL-E, and the Python standard REPL shell.

In the Foundation Engine tab, users can create a foundation model engine and config it to use different model parameters, such as temperature, maximum length, Top P, frequency penalty, presence penalty. In this way, users can create different engine instances of the same foundation model. Clicking ``Save Engine'' in the toolbar will save the created engine in the FM Engine tab, which can be edited later or exported to the projects for using.

In the Engines tab of an AI chain project, the user clicks the ``Import Engine...'' button, which will open the Engine Management view. The user can then drag the engines that the project will use to the engine editor. Then, clicking the ``Export to Project'' button will export the selected engines to the project. The user can change the configuration of the engines used in the project in the project's Engine tab, but these changes will not affect the original engines in the Engine Management.
The user can download the engine information to local files or upload it from local files to the IDE.

\section{User Study for Sapper}
\subsection{Design of Formative Study}
Our Sapper platform has the potential to significantly reduce the barriers to developing and utilizing state-of-the-art deep learning and large language models by offering a block-based visual programming technique, embedding a systematic prompting engineering methodology. 
\cy{In this section, we will assess the effectiveness of our tool (called Sapper V2) compared to the native programming tool, Python, which is currently the second most popular programming language used in GitHub repository (after JavaScript)\footnote{\cy{\url{https://octoverse.github.com/2022/top-programming-languages}}}.
Additionally, Python's reputation for being relatively easy to learn and use makes it a suitable choice for comparison. We will also evaluate an ablation version of our tool (called Sapper V1 - Block View Only). Our main objectives are to evaluate the ease of use and the learning curve associated with Sapper.}

\subsubsection{Procedure}
We recruited \cy{18} participants for the study, of which all aged 18-25 years. Seven participants major in Computer Science, three major in Software Engineering, and two major in Artificial Intelligence. Of the \cy{18} participants, \cy{10} participants were novice programmers (0-1 year of experience), \cy{six} were beginners (1-3 years of experience), and two were experienced programmers (3+ years of experience). \red{In terms of gender, there were 10 male and 8 female participants.} None of them have used visual programming tools before, and all have learned Python and use PyCharm before.

We conducted a within-subject study, in which each participant was asked to use three tools, native \textbf{Python} (\red{PyCharm}), \textbf{Sapper with Design View (Sapper V2)}, and \textbf{Sapper without Design View (Sapper V1)}, to program tasks in counterbalanced order. \cy{Counterbalanced order was employed to evenly distribute the potential sequence effects or learning biases associated with using different programming tools across participants, thereby minimizing the influence of order-related biases~\cite{cook2002experimental}}. 
In other words, since we had three different tools to choose from, there were a total of six possible combinations of orders in which the tools could be used. Each combination of orders was used by \cy{three} participants in the study.

Before starting the formal tasks, we provided a training session to introduce how to use Sapper with Design View. We also gave them some warm-up tasks to practice and ensure that they understand the basic usage. During the training session, participants were encouraged to ask any questions they had about Sapper. In addition, we also ask them to program the warm-up tasks using Python to be fair. 

Next, each participant was asked to use all three tools to program tasks in counterbalance order. We prepare three sets of tasks (Task A/B/C) with similar difficulties and each subtask in each set will assess the same programming constructs (i.e., plain use of LLM, if-else, while loop, variables \& use of a different LLM). For each task, we provided a written document with the overall goal and general approach. We specify the type of LLM to use, and provide test cases for them to verify their programs.

\cy{Taking task A as example, where participants were tasked with developing a service for a question-answering robot. 
They need to design a simple system where users can pose questions. The system, powered by the text-davinci-003 model, would then provide the answers (plain use of LLM). 
Task A2 added the capability to distinguish between code-related and general questions, using distinct prompts for each (if-else). 
In Task A3, the system could continue answering questions if the participants wished (while loop). 
Finally, in Task A4, a variable is used to maintain the conversation history. After the participants ends the conversation, the participants need to use GPT-3.5-turbo model to summary all questions asked based on the history (variables \& use of a different LLM). 
Task B/C, similar in complexity to Task A, involved the same programming constructs for each subtask. Further details regarding all tasks can be found on our GitHub repository\footnote{\url{https://github.com/YuCheng1106/PromptSapper}}.
}

Doing this can reduce the potential bias introduced by irrelevant factors, such as one may include a testing phase while another may not. Once participants finished all tasks using all tools, they were asked to fill in a questionnaire regarding user experience and the cognitive dimensions for interface evaluation~\cite{blackwell2000cognitive}. During the programming process, the computer screen was recorded to assist following analysis\footnote{We obtained their approval for this}.

We utilize paired t-tests to assess the statistical differences among three tools. Given that our evaluation involves multiple tools (n=3), we will implement the Benjamini-Hochberg method~\cite{benjamini1995controlling} to adjust p-values for multiple inferential statistical tests. This approach helps in controlling the false discovery rate, ensuring more reliable conclusions. All p-values reported in our findings will be those that have undergone this correction process.

\begin{figure}
    \centering
    \includegraphics[width=0.8\textwidth]{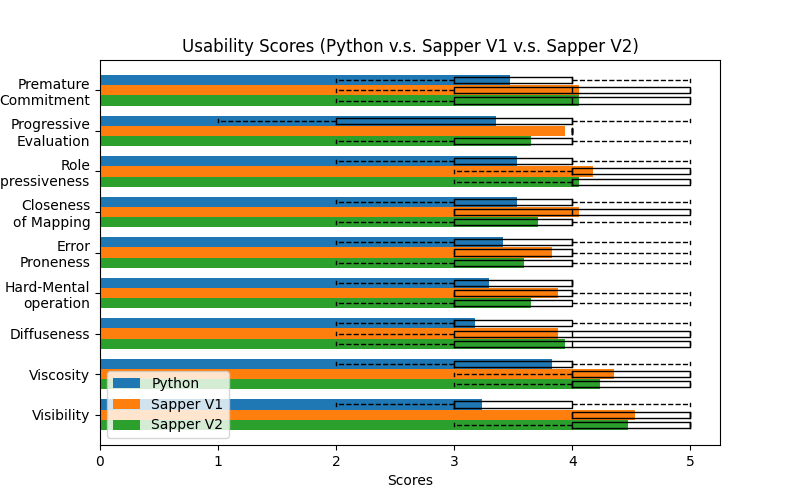}
    \caption{\cy{Usability scores for native Python, Sapper V1 and Sapper V2.}}
    \label{fig:usability_scores}
\end{figure}

\subsection{Results}

In this section, we report the results of our user study. We first compare Sapper V2 with Python, to evaluate the comprehensibility and the level of ease of use, and then compare Sapper V2 with Sapper V1, to understand the pros and cons of Design View.

\subsubsection{Sapper V2 VS. Python}

\textbf{Sapper is easier to understand and use:}
\label{sec:python_v2_time_correctness}
We first assessed the time spent and correctness of both tools by measuring the total time taken to complete Tasks 1-4 and the number of correctly completed tasks. We performed paired t-tests to determine if there were any significant differences in time spent and correctness between the two tools. The null hypothesis for both tests was that the two tools have similar distributions.

For correctness, there is no significant difference between Sapper V2 and Python, with a t-statistic of \cy{-0.53} and a p-value of \chyu{0.59} (>0.05).
For time spent, the result indicated a significant difference in time spent using the two tools, with a t-statistic of \cy{4.54} and a p-value of \chyu{0.0004} (<0.05). This strongly rejects the null hypothesis and shows that participants took significantly less time to complete the tasks using Sapper V2 (Mean=\cy{1,689.00}s, Std=\cy{447.49}s) than Python (Mean=\cy{2,366.00}s, Std=\cy{536.70}s). 
We carefully examined the videos and revealed two main observations. 
First, we found that Python is more error-prone than Sapper V2. For instance, when using Python, participants made simple grammar errors such as forgetting to convert input into an integer or initiate a variable, regardless of their programming expertise. 
In comparison, Sapper V2 can save the time spent correcting these errors. 
Second, participants struggled to locate the correct LLM API in Python and spent a considerable amount of time reading documentation to find a proper API call format when using Python. This issue was exacerbated when multiple APIs were used in different formats, leading to confusion and further time consumption. For example, while participants could rely on the same package ``openai'' to invoke GPT3.5 and GPT3 models, the parameter names for input prompt changed from ``prompt'' in GPT3 to ``messages'' in GPT3.5, causing confusion among participants.
In contrast, Sapper V2 encapsulates these complex API calls within a simple and integrated interface, reducing the time needed to locate and understand the correct API format. 
These observations show that our Sapper can provide more support to users by saving their time in learning programming language, debugging the syntax error, and instead can focus more on the logic and the functionalities.

\textbf{Usefulness Scores:}
To assess the usability of Python and Sapper V2, we analyzed the results in terms of 9 different cognitive dimensions~\cite{blackwell2000cognitive}, as depicted in Figure~\ref{fig:usability_scores}. The bar chart shows that Sapper V2 outperforms Python in terms of usability for all metrics, with higher scores indicating better usability.

We conducted a paired t-test for each metric and found that, \textit{except for diffuseness and visibility}, no significant differences were observed between Python and Sapper V2 (with $p > 0.05$) in all other metrics. 
Diffuseness refers to the conciseness of the notation used and how much space it requires to describe a task. A higher diffuseness score implies that the tool is more concise and requires less space to describe a task. Our statistical analysis (with $p = \chyu{0.0151} < 0.05$) revealed that Sapper V2 is significantly more concise and easier to use for participants, thus saving time and effort.
Visibility is a measure of how easily users can see and find the different parts of the notation while creating or changing it. It also includes the ability to compare or combine different parts simultaneously. A higher visibility score indicates that the tool is easier to modify and create new components. The result (with $p = \chyu{0.0006} < 0.05$) suggests that Sapper V2 is particularly convenient when making changes to the code due to its high visibility score.
These findings are also consistent with our previous observations that Sapper V2 simplifies the API calls and provides a more user-friendly interface, reducing the cognitive load on participants. Overall, our study suggests that Sapper V2 is a more usable tool.



\begin{table}
    \centering
    \begin{tabular}{l|c|c|c}
       \hline
               &  Python  & Sapper V1 & Sapper V2 \\
       \hline
       Task 1  & 475.27 (228.58) & 211.33 (76.54)  & \textbf{155.72 (49.65)} \\
       Task 2  & 574.50 (315.35) & \textbf{485.72 (178.33)} & 529.94 (214.53) \\
       Task 3  & 508.55 (195.43) & 382.05 (144.18) & \textbf{359.77 (130.30)} \\
       Task 4  & 807.66 (337.61) & 672.11 (338.02) & \textbf{644.11 (354.02)} \\
       \hline
       ALL     & 2,366.00 (536.70) & \textbf{1,751.00 (493.34)} & \textbf{1,689.00 (447.49)} \\
       \hline
    \end{tabular}
    \caption{Detailed time spent (second) for each task.}
    \label{tab:detailed_time_spent}
\end{table}

\subsubsection{Design View: Sapper V1 v.s. Sapper V2}
Similarly, we conducted paired t-tests to evaluate correctness, the time spent and usability scores between Sapper V1 and Sapper V2 to understand the pros and cons of design view.

In terms of correctness, no significant difference was found between Sapper V1 and Sapper V2.
No significance was found between Sapper V1 (Mean=\cy{1,751.00}s, Std=\cy{493.34}s) and Sapper V2 (Mean=\cy{1,689.00}s, Std=\cy{447.49}s) in time spent ($p=\chyu{0.62}>0.05$).
We carefully examine the usage videos from the participants.
While Sapper V2 can quickly generate some basic blocks for simple tasks, its current situation also brings some new issues to the participants. For example, it currently does not support copy and paste so that the participants have to create similar blocks repeatedly. In comparison, V1 supports the copy-paste function and the participants can duplicate existing blocks and only need to slightly revise some prompts or models.
Secondly, the current design view does not support automated generation of if-else or while loops, the participants should instead create one and perform draggings to put the related blocks under the condition or loops, which is similar to the block view in V1.
Third, our task decomposition copilot requires a significant time to get the response causing extra time consumption.
In addition to these, we surprisingly find that while the original purpose of the design view is for generating the basic blocks by given the task description, and then the participants can edit them in Block View, many participants instead directly use the design view to make changes, such as adding control flow in design view. 
However, as there are no significant time consumption differences observed, the current design view status can be used as an alternative to Sapper V1, and the users can choose the ones they prefer or use them interactively, which brings more flexibility and freedom to end-users.


For the usability scores, as seen in Figure~\ref{fig:usability_scores}, Sapper V1 obtains higher mean scores in most metrics except for premature commitment, diffuseness and viscosity. However, no significant statistic differences were found between Sapper V1 and Sapper V2 in each dimension  (with all $p > 0.05$), which indicates that these two variances of Sapper have similar user experience.
The reason may be that the current design view is also another no-code interface that allows participants to generate basic blocks and add if-loop, while loop, which requires no coding background. 

In terms of general usage of Sapper, we found that our Sapper provides great flexibility to users.
In Block View, some participants adopt the hierarchy workflow, while some use a vertical workflow with a variable to record the value and feed it as input to another worker. 
In Design View, some use the design view to directly generate basic code given the requirement and then modify it; while some directly build up the structure manually in the design view.



\cy{\section{User Study on The Effectiveness of Copilots}}
\label{sec:userStudyCopilots}
\cy{In the design view of Sapper IDE, we introduced two LLM-based co-pilots to assist users in requirement elicitation and AI chain skeleton generation, with the aim of supporting the primary activities during the design phase. In this section, we further assess the effectiveness of these two co-pilots.}

\cy{\subsection{Procedure}}
\cy{We recruited another 12 participants aged between 18 and 25 through public channels. All participants were local university students, with 6 specializing in Artificial Intelligence, 4 in Computer Science, and 2 in Big Data. Among these 12 participants, 4 had 0-1 years of development experience, 5 had 1-3 years of experience, and 3 had over 3 years of development experience.}

\cy{Before conducting the formal study, we provided training to the participants, which included familiarizing them with the basic operations and interface of Sapper IDE to ensure they had the necessary software operating skills. 
Additionally, we offered background knowledge on requirement elicitation and AI chain skeleton generation to help participants better understand and utilize the co-pilots, thus ensuring the credibility and accuracy of the experiment.}

\cy{\subsubsection{Requirement Elicitation Copilot's Study Procedure}
In the first part of the study, we focused on evaluating the effectiveness of the Requirement Elicitation Copilot. Each participant was presented with a abstract task requirement: ``I want to develop a service to help users prepare for interviews.'' Their task was to utilize our tool for requirement elicitation within 10 minutes.
Following this, participants were asked to think another task by themselves and clarify the requirement with our copilot within another 10-minute period. 
To gather valuable insights into their experiences, we administered a questionnaire that assessed user experience and interface evaluation dimensions.
Specifically, we asked the participants to rate usefulness, serendipity, diversity, ease-of-use and relevance by using 5-point Likert scale (1 being the lowest and 5 being the highest).  
Usefulness refers to how effectively our Copilot assists participants in clarifying their requirements. It assesses whether participants feel more capable of understanding their needs after using our Copilot. 
Serendipity evaluates the unexpected discoveries or insights participants may come across while using our copilot, going beyond their initial expectations. 
Diversity assesses the variety of insights and recommendations provided by the Copilot during requirement clarification. 
Ease-of-use examines the Copilot's user-friendliness, ensuring an efficient user experience. 
Relevance emphasizes the alignment of the Copilot's queries and suggestions with participants' specific requirements.}


\cy{\subsubsection{AI Chain Skeleton Generation Co-pilot's Study Procedure}
In the second part of the study, we aim to evaluate the effectiveness of the AI Chain Skeleton Generation Co-pilot.
Each participant was tasked with using the task descriptions they had previously conceptualized and obtained through requirement elicitation in the earlier study to execute AI chain skeleton generation copilot and run this AI chain skeleton.
Subsequently, they were asked to fill out another questionnaire to review the \textit{overall structure}, \textit{completeness of content}, and \textit{ease of modification} of the decomposed AI chain skeleton using 5-point Likert scale (1 being the lowest and 5 being the highest). 
The evaluation of the overall structure examines whether the task decomposition provided a clear and coherent framework for participants to understand the task effectively. 
Completeness assesses whether all the task details, including sub-tasks and steps, were adequately considered during decomposition to prevent any oversights. 
Ease-of-modification gauges participants' perception of how easily they could make adjustments to the decomposed structure, such as adding, removing, or editing sub-tasks or steps, to accommodate changes or optimize the task structure.
}

\cy{\subsection{Results}}
\cy{In this section, we report the evaluation results of the requirement elicitation and AI chain skeleton generation co-pilots. These findings offer valuable insights into the effectiveness of these two critical functionalities and the user experience, providing valuable insights for the design and future improvements of the Sapper IDE.}

\cy{\subsubsection{Results on Requirement Elicitation Co-pilot}
The 12 participants perform a total of 134 rounds of effective clarification question-answer sessions across the two requirement elicitation tasks (one provided by us and one self-conceived by participants). Among the 24 participant-task sessions, 2 sessions involve 4 rounds of clarification question-answer, 10 sessions involve 5 rounds, 8 sessions involve 6 rounds, and 4 sessions involve 7 rounds. The rounds of requirement elicitation question-answer are reasonable, considering the relatively short experimental time for each task, as well as the complexity of the requirement elicitation task.}





\cy{In the 12 self-conceived requirement elicitation tasks by participants, various domains were covered, including mental health, diet, health, finance, career planning, academics, and creativity. Specifically, these task categories encompass emotional and mental health (3 tasks), financial planning and career development tasks (3 tasks), cooking and nutrition (2 tasks), health and fitness (2 tasks), academic guidance (1 task) and creative writing (1 task). More details about the tasks and the participant requirement elicitation process can be found in our GitHub repository\footnote{\href{https://github.com/YuCheng1106/PromptSapper}{https://github.com/YuCheng1106/PromptSapper}}.}

\begin{figure}
    \centering
    \includegraphics[width=1\textwidth]{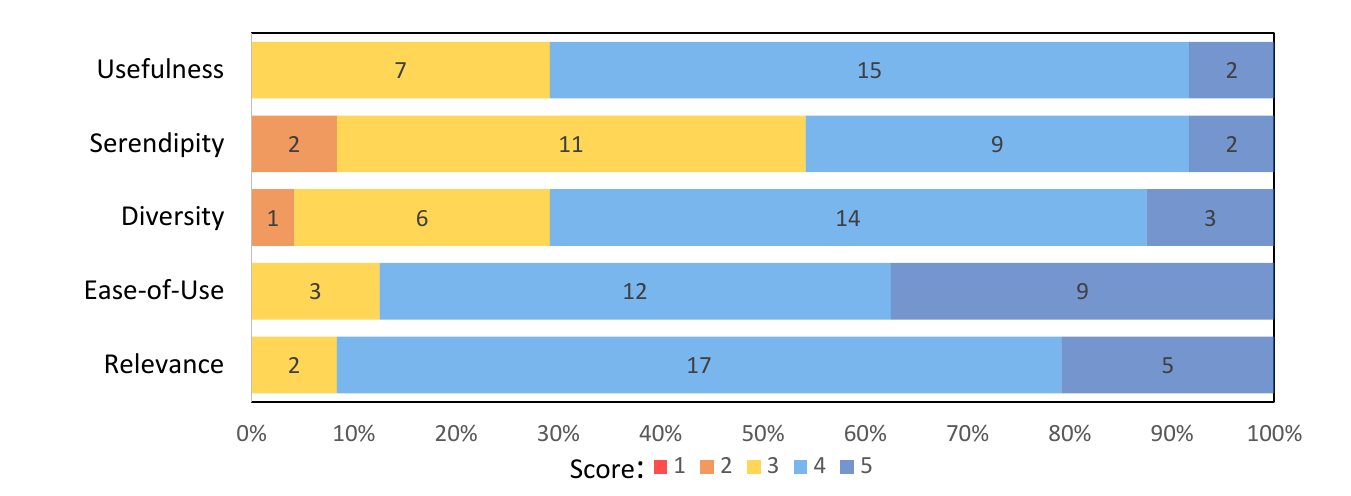}
\caption{\cy{Usability scores for requirement elicitation Co-pilot.}
    }
    \label{fig: score requirement}
\end{figure}

\cy{Figure \ref{fig: score requirement} presents the evaluation results of the five aspects' ratings for the requirement elicitation results by the 12 participants. 
For all these five aspects, the mean ratings are all larger or equal to 3.5, and the majority of scores fall in the range of 3 to 5. 
This indicates that our requirement elicitation copilot is satisfactory, helpful, easy to use, and provide diverse, unexpected and relevant suggestions.
In detail, among all 24 requirement elicitation sessions they evaluated, participants rated the usefulness and diversity of the requirement elicitation results with scores of 4 or 5 in over 70\% of the sessions. Additionally, in over 80\% of the ratings, participants rated the ease-of-use and relevance of the requirement elicitation results with scores of 4 or 5. Moreover, for 11 of the ratings (45.83\% of the total), participants rated the serendipity of the requirement elicitation results at 4 or 5.
}

\cy{We carefully checked the 24 requirement elicitation processes and find that our Copilot not only helped in clarifying and deepening their thoughts but also inspired them. 
For instance, when the participants use our tool to elicit the requirement for the mock interview tool, it first confirmed the objective, offering enhanced detail such as ``an interview preparation service that provides interview questions and offers tips.''
Furthermore, to spark inspiration, the co-pilot posed questions about potential features, suggesting, ``Let's consider how your service might evaluate a user’s performance.''
Importantly, it refrains from repetitively asking similar questions within a single session. This effectiveness is attributed to our design, where the conversation histories are transmitted to the LLM during the whole session. Moreover, our chat-style interface is well-suited for a conversational question-answer context. Consequently, our Requirement Elicitation Copilot receives high ratings in these five aspects.}

\begin{figure}
    \centering
    \includegraphics[width=1\textwidth]{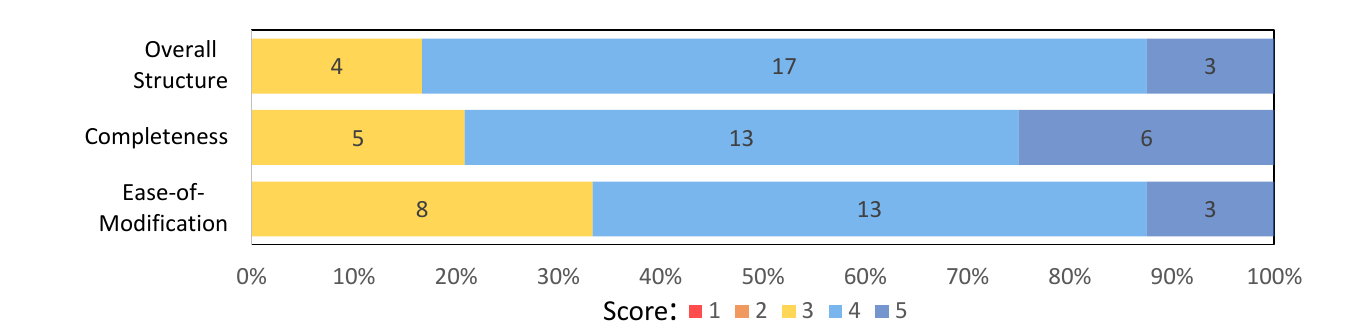}
    \caption{\cy{Usability scores for AI chain skeleton generation Co-pilot.}}
    \label{fig: score generation}
\end{figure}

\cy{\subsubsection{Results on AI Chain Skeleton Generation Co-pilot}
As shown in Figure \ref{fig: score generation}, among all 24 tasks, all participants assigned scores of 3 or higher for all three metrics, with most ratings falling between 4 and 5. 
This indicates that our Copilot effectively assists users in generating AI chain skeletons in a complete and correct manner. }
\cy{We carefully reviewed and executed the AI chain skeletons generated by the 12 participants. Out of these, 18 could be directly executed to obtain results, and for the remaining six, minor modifications to the generated prompts were sufficient to achieve the desired outcomes. Upon inquiry, most participants mentioned that while the generated skeletons were executable and provided results, some adjustments were necessary. 
These adjustments might involve adding additional steps to refine the functionality or fine-tuning the prompts to align the service output more closely with their expectations. Nonetheless, they all expressed that the generated AI skeletons greatly facilitated their subsequent development efforts.}

\section{Discussion}
The emerging AI chain paradigm is rapidly evolving, and we envision and discuss several issues that are important to the development and practical application of the AI chain paradigm.

\textbf{Enhancement of AI Chain Methodology and IDE:}
Based on extensive literature, community shared experiences, and our own experiences with foundation models, we have presented the first methodology and IDE to support the AI chain paradigm. 
We look forward to refining this methodology and IDE based on feedback from practical adoption of the proposed methodology and the current IDE. 
For example, we plan to enhance the co-pilots in the design view to enable the generation of condition and loop skeleton. 
We also plan to develop a prompt debugging co-pilot to better support interactive prompt debugging. 
Therefore, at the methodology level, an important question is how to integrate the AI chain methodology into existing traditional software DevOps and machine learning MLOps practices. 
As the core philosophy of the proposed AI chain methodology is rapid prototyping, this does not fundamentally conflict with existing DevOps and MLOps, but the integration may require adjustments at both sides, such as how to combine human-centric AI chain development with automatic continuous integration/delivery.


\textbf{Testing Generative Models:}
AI chain testing is currently ad hoc and faces many challenges. 
First, the core of AI chain testing is prompt testing, and the design space of prompts is vast. 
Besides the flexibility of natural language, many prompt design choices can influence the performance of a prompt. 
Prompt testing is currently subjective, and there is no systematic method for it. 
Second, since the emergent ability of generative models is customized through in-context learning, are we facing a similar Schrodinger's cat problem in quantum physics? 
If observing and measuring have a fundamental impact on the observed object, can we still fully test the essence of generative models, or can we only test the appearances of a specific customized version by in-context learning?
Last but not least, the fast iteration of foundation models can cause the moving-sand challenge. 
This can cause difficulties in maintaining the stability and effectiveness of the AI chain, especially when the changes in the foundation models affect the behavior of the workers and their interaction with AI. 
These changes require timely testing and adjustments to the AI chain and its component workers.
All these challenges require new research and innovation in testing generative models and AI chains on top of them.

\textbf{Future Deployment:}
Our AI chain IDE directly serves AI chain engineers (although referred to as engineers, they can be anyone who wants to develop AI chains on top of foundation models). 
The AI services developed by these individuals need to reach end users through various platforms, such as social media, office software, development tools, scientific platforms. 
Our current IDE has demonstrated the feasibility of this development and deployment approach, particularly in the form of chatbot services. 
In addition to chatbot services, we plan to seamlessly integrate AI chain-based services into people's daily workflows, providing customized work co-pilots. 
In addition to AI service integration, it will also be important to integrate AI chain development with other design, prototyping, and development tools such as Figma~\cite{figma}, Jupyter notebook~\cite{Jupyter}, and Replit~\cite{replit}. 
For example, integrating AI chain development into Figma could allow UI/UX designers to quickly design prototypes with actual AI functionality, rather than simple prefabricated mockups. Similarly, integrating AI chain development into Jupyter notebook could enable data scientists to more conveniently utilize foundation model services in their data/ML projects.

\red{\section{Data and Code Availability}
We released the source code and experimental data in our GitHub Repository~\footnote{https://github.com/YuCheng1106/PromptSapper}.
We also release the prototype of our Prompt Sapper and the link can be found in the Repository.
}
\section{Conclusion}
This paper presents a systematic methodology for AI chain engineering, encompassing three key concepts and four activities to advance the modularity, composability, debuggability and reusability of AI chain functionalities. 
By incorporating this methodology, we introduce Prompt Sapper, a blocky-style visual programming tool that empowers AI chain engineers, including non-technical users, to compose prompt-based AI services using foundation models through chat-based requirement analysis and visual programming. 
Sapper contains two views, namely Design View and Block View, with LLM co-pilots, to help elicit requirements, generate code skeleton, run/test the AI service.
In a user study involving \cy{18} participants, we demonstrate the low entry barrier and practicality of Prompt Sapper, showcasing its significant time-saving capabilities compared to traditional IDEs. The tool eliminates the need to learn various API calls and minimizes time spent on resource searching. \cy{Moreover, through a separate user study with 12 participants, we confirmed the effectiveness of our LLM-based co-pilots.}

\begin{acks}
This work is partially supported by the National Nature Science Foundation of China under Grant (62262031).
\end{acks}

\bibliographystyle{ACM-Reference-Format}
\bibliography{reference}

\end{document}